\documentclass[twoside]{jpconf}

\usepackage{blindtext}
\usepackage[overload]{textcase}
\usepackage{threeparttable}

\def\asec{\ifmmode ^{\prime\prime}\else$^{\prime\prime}$\fi}

\def\it{\sl}
\def\degs{\ifmmode ^{\circ}\else$^{\circ}$\fi}
\def\amin{\ifmmode ^{\prime}\else$^{\prime}$\fi}
\def\asec{\ifmmode ^{\prime\prime}\else$^{\prime\prime}$\fi}

\def\farcs{\hbox{$.\!\!^{\prime\prime}$}}

\def\degs{\ifmmode ^{\circ}\else$^{\circ}$\fi}
\def\amin{\ifmmode ^{\prime}\else$^{\prime}$\fi}
\def\farcm{\hbox{$.\mkern-4mu^\prime$}}
\def\eqalign#1{\null\,\vcenter{\openup1\jot \m@th
   \ialign{\strut\hfil$\displaystyle{##}$&$\displaystyle{{}##}$\hfil
   \crcr#1\crcr}}\,}
\def\j1045{J1045-4509}
\def\psr{J1045}

\usepackage{graphicx}
\usepackage{mathtools}
\usepackage{enumitem,amssymb}
\newlist{todolist}{itemize}{2}
\setlist[todolist]{label=$\square$}
\usepackage{pifont}

\usepackage{subfigure}
\usepackage{hyperref}

\begin{document}

\title{Optical identification of binary system millisecond pulsar J1045$-$4509 with the VLT}

\author{A V Bobakov$^1$, D A Zyuzin$^1$ and Yu A Shibanov $^1$}

\address{$^1$Ioffe Institute, Politekhnicheskaya 26, St. Petersburg, 194021, Russia}

\ead{bobakov\_alex@mail.ru}

\begin{abstract}We analyse archival optical data on the binary companion to the millisecond pulsar \j1045\ obtained with the Very Large Telescope. 
A possible optical counterpart candidate is found at the pulsar position.
Its magnitudes are $V>26.4$, $R=25.7(2)$ and $I=25.4(2)$.  
The data are compared with  white dwarf evolutionary models. Depending on still poorly known distance to the pulsar, two alternative interpretations are possible. 
 For the radio timing parallax distance, the candidate can be a very cool and old  
white dwarf with the hydrogen atmosphere and the temperature of $< 3000$ K   and  the age of $> 5$ Gyr. 
 The dispersion measure distance suggests  a younger ($\sim$ 2 Gyr) and hotter ($\sim$ 6000 K) white dwarf, whose atmosphere composition remains unknown.

\end{abstract}

\section{Introduction}

From about 2700 pulsars known (ATNF pulsar catalogue\cite{2005yCat.7245....0M}), 
$\sim$13\%
are millisecond pulsars (MSPs) which 
have short periods ($P < $ 30 ms) and relatively low magnetic fields ($B\sim 10^8-10^{10}$ G).  
MSPs are considered to be neutron stars (NSs) spun-up (or `recycled')  
by accretion of matter from main-sequence companions \cite{Bisnovatyi-Kogan1974}.
Most of them ($\sim$80\%) are members of binary systems. 
Depending on  initial conditions, the companion of a MSP can be a helium/carbon white dwarf (WD),
a main sequence star, a non-degenerate or partially degenerate stellar core  
or a neutron star \cite{manchester2017}.
Multiband optical photometric and spectroscopic observations 
play a crucial role in studying binary MSPs.
They allow one to constrain the companion star properties (its spectral class, mass and age),
to determine the distance to a particular system and to independently constrain the binary 
and pulsar parameters 

when they are poorly defined from the radio data. 
Revealing the nature of the companion  allows us to test the existing theories 
of MSPs binary evolution.

The binary MSP J1045$-$4509 (hereafter J1045) was discovered in the radio 
with the Parkes telescope in 1993 \cite{1994ApJ...425L..41B}.
Its parameters are presented in table~\ref{parameters}. 
The distance to the pulsar was estimated 
from the dispersion measure using Galactic electron maps NE2001  \cite{2002astro.ph..7156C} and YMW16\cite{ymw} 

and from the radio timing parallax \cite{2016MNRAS.455.1751R}.
We used archival optical data 
from the Very Large Telescope (VLT) 
to  identify the binary companion of this pulsar and to study its properties. 

 \begin{table}[h]
         \caption{Parameters of the binary MSP J1045-4509  \cite{2016MNRAS.455.1751R}} 
     \begin{center}
     \begin{tabular}{lc}
     \br
         Right ascension $\alpha$(J2000) & 10:45:50.18696(3) \\
         Declination $\delta$(J2000) & $-$45:09:54.1223(4)\\
         Galactic longitude \textit{l} (deg) & 280.85\\
         Galactic latitude \textit{b} (deg) & 12.25\\
         Epoch (MJD) & 54500\\
         Proper motion $\mu_{\alpha}=\Dot{\alpha}\cos{(\delta)}$ (mas yr$^{-1}$) & $-$6.07(9)\\
         Proper motion $\mu_{\delta}$ (mas yr$^{-1}$) & 5.20(10)\\
         \mr
         Spin period P (ms) & 7.47(8)\\
         Period derivation $\Dot{P}$ ($10^{-20}$~s s$^{-1}$) & 1.7662(5)\\
         Characteristic age (Gyr) & 6.7\\
         Spin-down luminosity $\Dot{E}$ (erg s$^{-1}$) & $1.7\times10^{33}$ \\
         \mr
         Orbital period $P_{\textit{b}}$ (days) & 4.08352925(3)\\
         Eccentricity ($10^{-6}$) & 23.67\\
         Companion min. mass$^{a}$
         & 0.159 $M_{\odot}$\\ 
         \mr
         Parallax $\pi$ (mas)& 2.2$\pm$1.1 \\
         Dispersion measure $DM$ (pc cm$^{-1}$)& 58.17\\
         Distance $D_{YMW16}$ (kpc) & 0.34\\
         Distance $D_{NE2001} $(kpc) &1.95 \\
         Distance $D_{\pi}$ (kpc) & $0.34^{+0.20}_{-0.10}$\\
         \mr
     \end{tabular}
     \begin{tablenotes}
\item $^a$The minimal companion mass was calculated assuming the orbit inclination angle $i=90$ deg 
and the pulsar mass $M_p=1.35$M$_\odot$.
\end{tablenotes}
          \end{center}
      \label{parameters}
 \end{table}

\section{Observations and data reduction}
\label{sec:data}
The field of J1045 was observed with the VLT on 2001 February 20 in three Bessel filters 
$V$, $R$ and
 $I$ using the FORS1 instrument. The FORS1 field of view and pixel scale were 
3\farcm4$\times$3\farcm4  and 0\farcs1, respectively.

Ten dithered images were obtained in each filter resulting in  total exposure times of 1400, 800, and 1400 s  
for the  $V$, $R$, and I filters, respectively.  
Standard data reduction and analysis were performed using  IRAF  packages.

Ten bright unsaturated stars were used for astrometric calibration. 
Their coordinates with uncertainties of about $\approx$ 0.7 mas 
were taken from the Gaia DR2 catalogue \cite{2016A&A...595A...1G}.
The resulting formal astrometric fit uncertainties in RA and Dec were 
 $\approx$0\farcs05. 

For the photometric calibration, we used the Landolt
standard stars SA 95-107 \cite{SA95-105} observed during the same night as the target. 
The derived photometric zeropoints are 
$Z_{V}=27.251(11)$, $Z_{R}= 27.180(20)$ and $Z_{I}=26.502(13)$.
 
We used the atmospheric extinction coefficients
$k_{V}=0.13(1)$, $k_{R}=0.075(10)$ and  $k_{I}=0.056(10)$ taken from VLT website\\
(\url{http://www.eso.org/observing/dfo/quality/FORS2/qc/photcoeff/photcoeffs_fors2.html}).

\section{Results}
\begin{figure}[]
 \centering
 \includegraphics[scale=0.20]{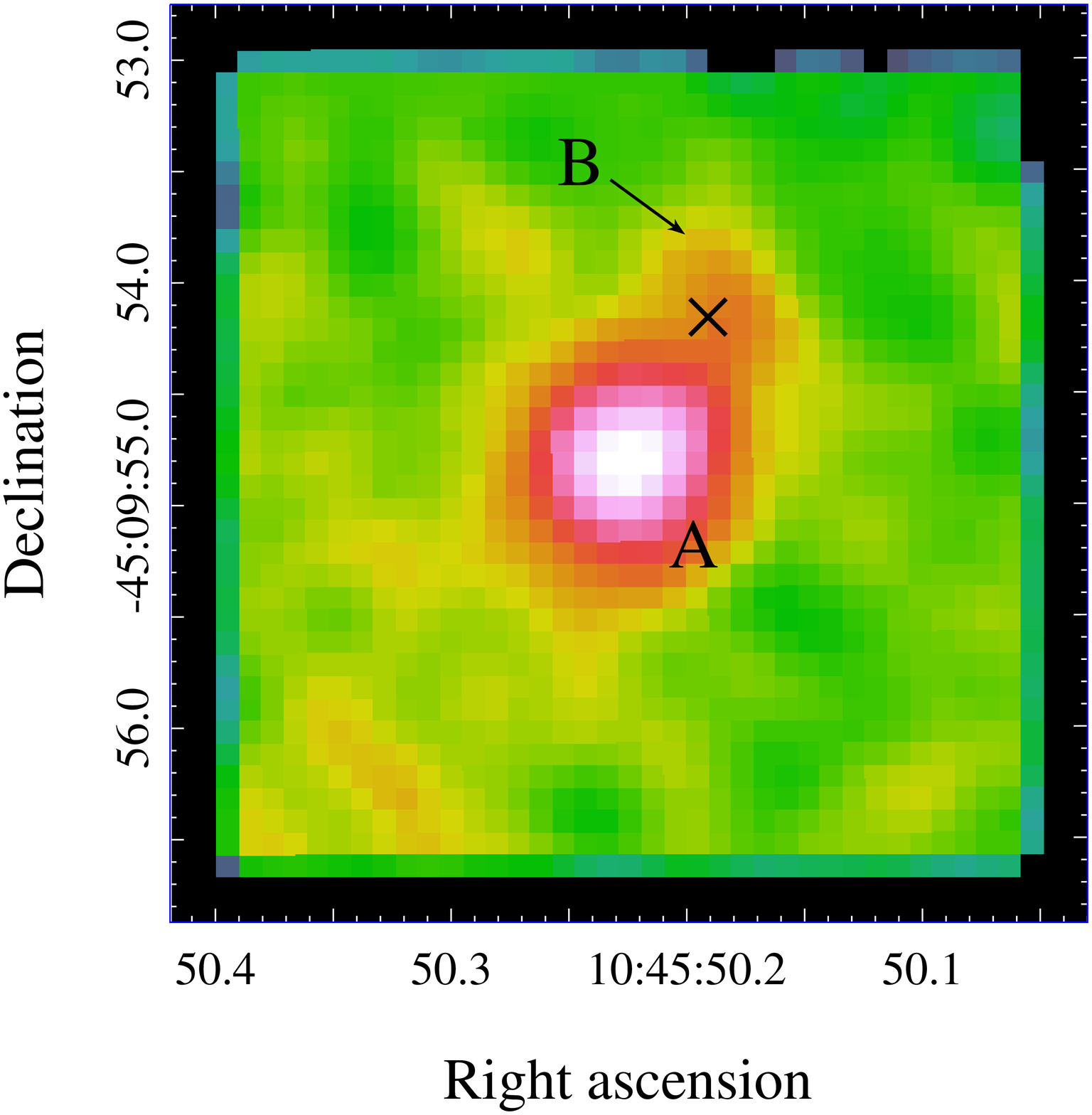}
 \includegraphics[scale=0.20]{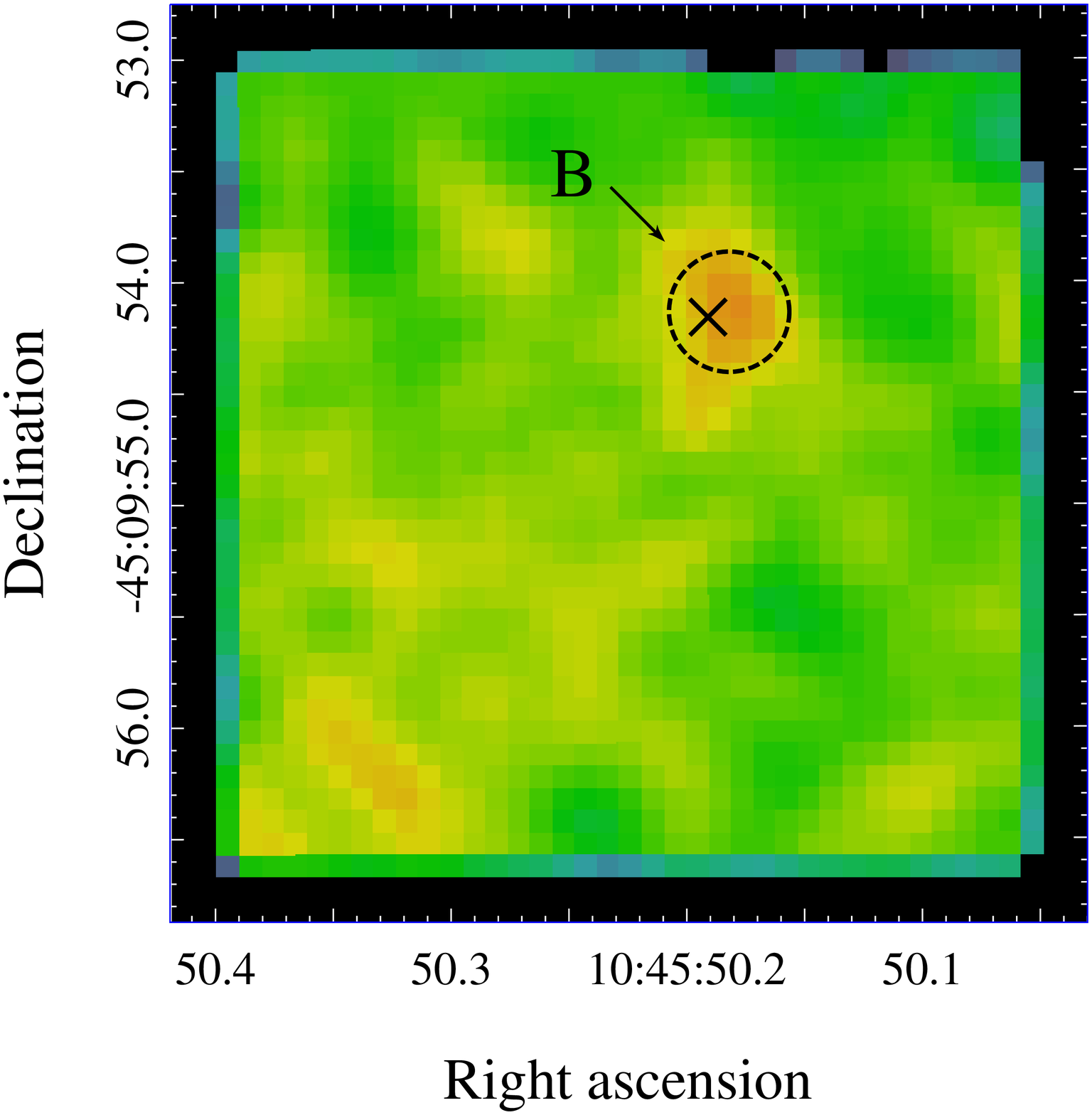}
\caption{4\asec$\times$4\asec\ VLT $I$-band images of the 
\psr\ field. In the left panel, the  
optical sources in the nearest vicinity of the pulsar are marked by ``A'' and ``B''.
In the right panel, the brighter source ``A'' was subtracted
to better reveal possible optical counterpart 
of the binary.
The solid circle with the radius of 0\farcs25 shows its 3$\sigma$ position uncertainty.
The pulsar radio position is marked by the ``X'' symbol.}
\label{fig:i-image}
\end{figure}

We have found possible optical counterpart to J1045 in $I$ and $R$ bands.
Its coordinates are RA=10:45:50.230 {and} Dec=$-$45:09:54.917. 
Within 3$\sigma $ uncertainty of 0\farcs25 it coincides with the pulsar radio 
position RA=10:45:50.191(9) and Dec=$-$45:09:54.160(10) corrected for the  proper motion (table \ref{parameters}) to the optical observation epoch. The image in  the $I$ band is presented in figure \ref{fig:i-image}.  
The counterpart candidate marked by ``B'' partially overlaps with a background source ``A''. We subtracted the source ``A'' using the IRAF point spread function (PSF) algorithm. This allowed us to do photometry of the source ``B''. The resulting candidate magnitudes are $V>26.4$, $R=25.7(2)$ and $I=25.4(2)$. 

    To correct these values for the interstellar extinction, we used the dust map \cite{dustmap} and distances to J1045 (table \ref{parameters}). The reddening $E(B-V)$ is 0.03(2) and 0.06(1) for the minimum and maximum distance estimates.  We used the absorption law calculator
    (\url{http://www.dougwelch.org/Acurve.html})
    to  determine the extinctions $A_{\lambda}$ in the observed bands. 
    
\begin{figure}[]
\centering
\includegraphics[scale=0.30]{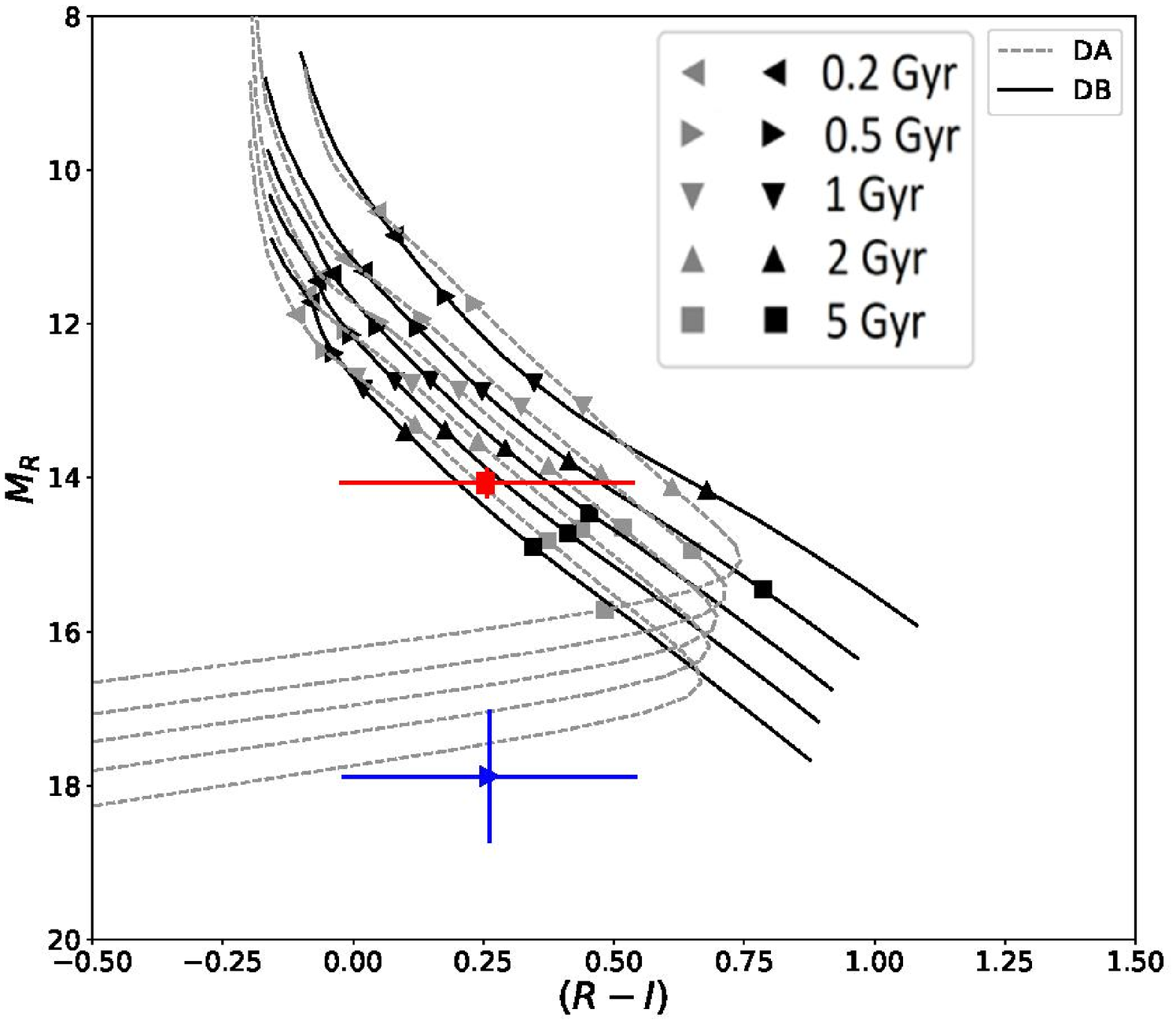}
\includegraphics[scale=0.30]{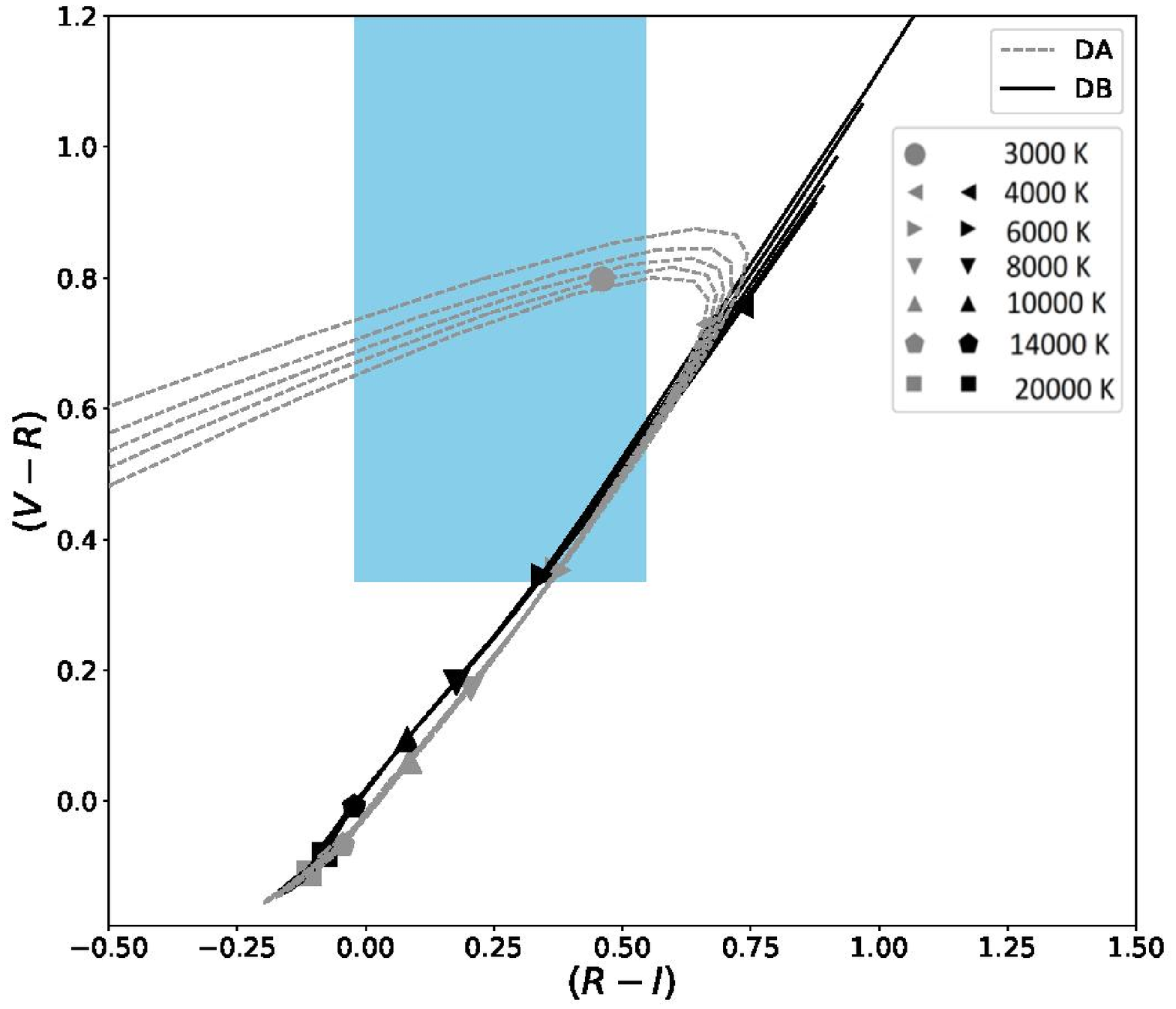}
\caption{
Colour-magnitude (left) and colour-colour (right) diagrams with various WD cooling tracks.  
Solid black (DA) and dashed gray (DB) lines show the cooling tracks for WDs 
with hydrogen and helium atmospheres, respectively 
\cite{2006AJ....132.1221H, 2006ApJ...651L.137K,2011ApJ...730..128T,2011ApJ...737...28B}, 
with masses 0.2--1~${\rm M_\odot}$ (spaced by $0.2~{\rm M_\odot}$) increasing from upper to lower curves.
Cooling ages in the left panel and temperatures in the right panel 
are indicated by different symbols.
The location of the \psr\ presumed companion 
in the left panel is marked by error bars, where the red one corresponds
to the 1.9 kpc distance while the blue one -- to the pulsar timing parallax 
distance of $0.34^{+0.20}_{-0.10}$ kpc.
In the right panel, the location range of  
the companion  is constrained by the
light-blue stripe.
}
\label{diagrams}
\end{figure}
  
For the distance $D_{NE2001}=1.95$ kpc, the resulting 
intrinsic colours and absolute magnitude are 
$(I-R)_0$= 0.3(2), 
$(V-R)_0$ $> $ 0.4 
and $M_{R}$= 14.1(2). For $D_{\pi}$= 0.34$^{+0.2}_{-0.1}$ kpc 
$M_{R}$= 17.9(9) while the colours within the errors remain the same.

We compared the derived colours and absolute magnitudes of the source
with WD cooling tracks taken from \cite{2006AJ....132.1221H, 2006ApJ...651L.137K,2011ApJ...730..128T,2011ApJ...737...28B} (figure~\ref{diagrams}). 
As seen from the colour-magnitude and colour-colour diagrams, the source can be  
very cold ($<$3000 K) and old ($>$5 Gyr) hydrogen-atmosphere WD.
 MSP companions with similar characteristics are rare. Only a few of them with  
 temperatures below 3000 K  are known (e.g., J2017$+$0603 and J1231$-$1411 \cite{bassa2016a}). 
For the D$_{NE2001}$ distance the companion can be a hotter, 
$\sim$ 6000 K,  and younger WD with the age of $\sim$ 2~Gyr, while its atmosphere composition remains unknown. 

More accurate determination of the  distance  to the binary system 
from radio observations is needed to reveal
the real nature of the optical counterpart  candidate.
Measurements of the candidate proper motion can be a direct evidence of its association with the pulsar.

\ack{Thanks anonymous referee for useful comments and suggestions. Based on observations collected at the European Organisation for Astronomical Research in the Southern Hemisphere under ESO programme 66.D-0274(C).  DAZ thanks Pirinem School of Theoretical Physics for hospitality.}

\section*{Bibliography}
\bibliography{Bibliograph}

\end{document}